\documentclass[a4paper]{natureML}
\usepackage{amsmath}
\usepackage{amssymb}
\usepackage[numbers,sort&compress,super]{natbib}
\usepackage{doi}
\usepackage{graphicx}
\usepackage{color}
\usepackage{siunitx} 
\usepackage{upgreek}
\usepackage{longtable}
\usepackage{array}
\usepackage[singlelinecheck=off]{caption}
\usepackage[retainorgcmds]{IEEEtrantools}
\usepackage{MLHMM}
\usepackage{xr}
\newcommand{\ML}[1]{#1}%\textcolor{red}{#1}}

\title{Simulated single molecule microscopy with SMeagol}  
\author{Martin Lind\'en, Vladimir \'{C}uri\'{c}, \ML{Alexis
    Boucharin}, David Fange, \& Johan Elf$^\dagger$\\ Department of Cell and
  Molecular Biology, Uppsala University, Sweden.
$^\dagger$johan.elf@icm.uu.se}

\begin{document}
\maketitle
\addcontentsline{toc}{chapter}{Main text}
\medskip\\
\spacing{1} % to force single spacing for readability
\textbf{Summary:} SMeagol is a software tool to simulate highly
realistic microscopy data based on spatial systems biology models, in
order to facilitate development, validation, and optimization of
advanced analysis methods for live cell single molecule microscopy
data.\\ 
\textbf{Availability and Implementation:} SMeagol runs on Matlab R2014
and later, and uses compiled binaries in C for reaction-diffusion
simulations. Documentation, source code, and binaries for recent
versions of Mac OS, Windows, and Ubuntu Linux can be downloaded from\\
\url{http://smeagol.sourceforge.net}.\\ 
\textbf{Supplementary information:} Supplementary data are available
at \textit{Bioinformatics} online.\medskip\\
Recent advances in single particle tracking (SPT)
microscopy\cite{manley2008} make it possible to obtain tens of
thousands macromolecular trajectories from within a living cell in
just a few minutes. Since molecules typically change their movement
properties upon interactions, these trajectories contain information
about both locations and rates of intracellular reactions. This
information is unfortunately obscured by physical limitations of the
optical microscope and noise in detection systems, making statistical
methods development for SPT analysis a very active research
field. Unbiased testing and comparison of such methods are however
difficult given the absence of in vivo data of intracellular dynamics
where the true states of interaction are known, a.k.a. the “ground
truth”. A common resort is to instead use simulated, synthetic,
data. However, tests using such data give unrealistically optimistic
results if the simplifying assumptions underlying the analysis method
are exactly satisfied.  The need for realistic simulations is long
recognized in microscopy and systems biology
\citep{fullerton2012,ursell2012,cox2012,chenouard2014,sinko2014,sage2015,slepchenko2002,kerr2008,takahashi2010,fange2012,andrews2012},
but systematic combinations of the two are only currently emerging
\citep{angiolini2015,watabe2015}.

%, a practice known as "inverse crimes" \citep{kaipio2006}.
\begin{figure*}[!b]
%\fboxsep=0pt\colorbox{gray}{
%  \begin{minipage}[t]{235pt} \vbox to 100pt{\vfill\hbox to
%      235pt{\hfill\fontsize{24pt}{24pt}\selectfont FPO\hfill}\vfill}
%\end{minipage}}
\centerline{\includegraphics{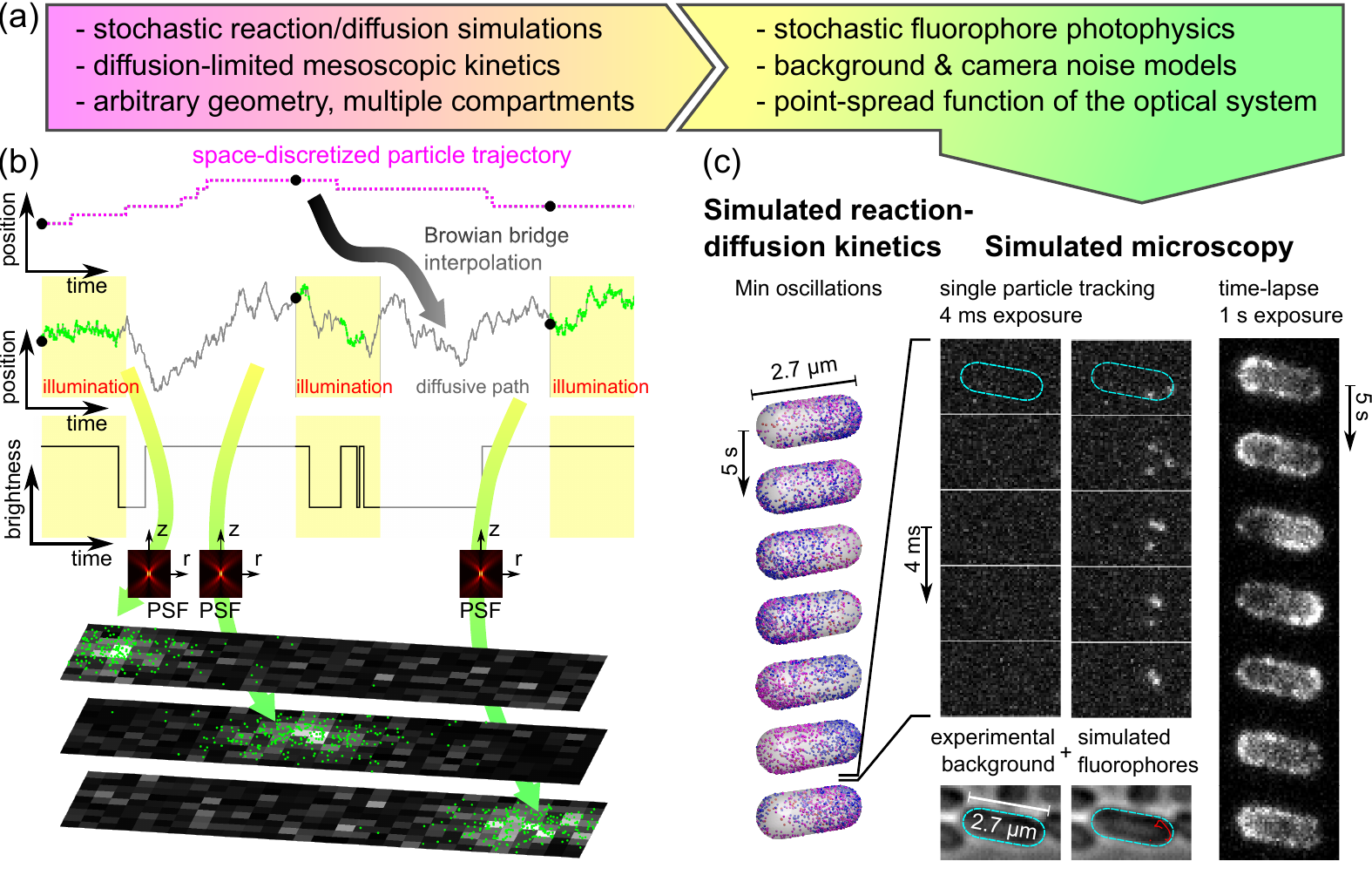}}
  \caption{\label{fig1} Simulated microscopy with SMeagol. (a)
  Workflow from stochastic reaction-diffusion simulations to
  images. (b) The microscopy simulation starts from trajectories
  generated by stochastic reaction-diffusion simulations, fills in
  stochastic motion and photon emission events between the trajectory
  points, and finally combines PSF and camera noise models to simulate
  realistic images. (c) Simulated microscopy of fluorescently labeled
  MinE proteins in the Min oscillatory system. Left: Stochastic
  reaction-diffusion simulation. Mid columns: Simulated SPT microscopy
  using an actual experimental background noise movie with continuous
  illumination and 4 ms/frame. Right: A simulation of continuous
  illumination and 1 s/frame renders a conventional (non-single
  molecule) fluorescence microscopy time-lapse movie. See also
  Supplementary movies S1,S2, and the Supplementary material for
  further details. }
\end{figure*}

We present the SMeagol package, that has been developed to generate
highly realistic single molecule microscopy time-lapse image series
aimed primarily at single particle tracking applications. The purpose
of SMeagol is to enable realistic comparisons between the output of
advanced analysis methods and known ground truth. SMeagol includes an
extended MesoRD \citep{fange2012} version for simulation of 3D
diffusion in cellular compartments, diffusion limited reaction
kinetics, surface adsorption, reactions in membranes, and other
complex aspects of reaction diffusion kinetics that do occur in cells,
but are not considered in SPT analysis algorithms. In addition to the
molecules' trajectories, SMeagol integrates the 3D point spread
function of the microscope, the kinetics of photo-activation, blinking
and bleaching of the simulated fluorophores, background noise, and
camera specific parameters (Figure 1\vphantom{\ref{fig1}}, movie
S1,S2). Great flexibility is allowed by the possibility to supply
these characteristic parameters either as tabulated experimental data
for a particular optical setup, or as theoretical models. The
combination of using reaction diffusion kinetics in cellular
geometries and physics-based simulations of the emission and detection
processes makes the images more realistic than the synthetic data used
for example by \citealp{chenouard2014}.

SMeagol can be used to optimize imaging conditions for specific
systems in silico and to benchmark methods for SPT analysis in analogy
with the methods that has been developed to benchmark localization
methods for non-moving single particles \citep{sage2015}.  
%For example, in the supplementary material we explore the robustness
%against localization errors and motional blur of the vbSPT software,
%which extracts multi-state diffusive models from SPT data
%\citep{persson2013}.
\ML{In the supplementary material, we explore the robustness against
localization errors and motional blur of the vbSPT software, which
extracts multi-state diffusive models from SPT data
\citep{persson2013}, and find that these effects can induce
overfitting under certain conditions. In addition, we provide a
  number of examples highlighting possibilities, limitation and
  computational requirements of the SMeagol simulation engine.}
%We also describe model examples to illustrate the computational
%requirements for large systems such as eucaryotic cells, modelling
%spatial heterogeneiety and non-diffusive transport processes, and
%construction of models with complex geometry.}

When combined with increasingly refined simulations of intracellular
processes, photo-physics and optics; live-cell microscopy is moving
closer to methods in fundamental physics, where combined simulation of
physical processes and detection systems have guided experimental
design and data analysis for a long time.
\vspace*{-10pt}

%\enlargethispage{6pt}
%\vadjust{\pagebreak}

\section*{Acknowledgements}
We thank Fredrik Persson and Elias Amselem for helpful discussions
about microscopy.
\vspace*{-12pt}
\section*{Funding}
This work was supported by the European Research Council,
Vetenskapsrådet, the Knut and Alice Wallenberg Foundation, the
Foundation for Strategic Research, and the Swedish strategic research
programme {eSSENCE}.
\vspace*{-12pt}

%\noindent
%{\bf Martin Lindén, Vladimir \'{C}uri\'{c}, \ML{Alexis Boucharin}, David
%  Fange, \& Johan Elf}

%\noindent
%Department of Cell and Molecular Biology, Uppsala University, Sweden.\\
%\noindent
%e-mail: johan.elf@icm.uu.se

%\input{acknowledgement.tex}

%\input{mainrefs.bbl}

\paragraph{References}
\addcontentsline{toc}{section}{References}
%\bibliographystyle{unsrtnat}
%\bibliography{references}

\newpage
%%%% % numbering w prefix
\renewcommand{\thesection}{S\arabic{section}}  
\renewcommand{\thetable}{S\arabic{table}}  
\renewcommand{\thefigure}{S\arabic{figure}}
\renewcommand{\theequation}{S\arabic{equation}}
%%%% 

\section*{Supplementary material}
\addcontentsline{toc}{chapter}{Supplementary material}
%A more complete SPT-time series bibliography
%\cite{calderon2013,calderon2014,persson2013,masson2009,masson2014,beheiry2015,berglund2010,michalet2012,vestergaard2014,monnier2012,monnier2015}.

%\begin{table}
%\caption{\label{tab:rd}Popular reaction-diffusion simulation packages.}
%  \begin{tabular}{ll}
%%    name & url & ref\\
%    \hline
%    MesoRD\cite{fange2012} $^\dagger$ & 
%    \url{http://mesord.sourceforge.net/}\\
%    GFRD\cite{takahashi2010} & \url{http://gfrd.org/}\\
%    MCell\cite{kerr2008}& \url{http://www.mcell.org/}\\
%    Smoldyn\cite{andrews2012}&\url{http://www.smoldyn.org/}\\
%    Vcell\cite{slepchenko2002}&\url{http://vcell.org/}\\
%    \hline
%  \end{tabular}\\
%{\small $^\dagger$ Capability for SMeagol-compatible output.}
%\end{table}

\section{Simulated microscopy with SMeagol}
SMeagol is a Matlab software suite that simulates microscopy images of
randomly moving particles using two main ingredients: diffusive motion
and stochastic photon emission events. In addition, noise from various
sources (camera, background, optics, blinking and photobleaching, etc.)
can be included in a modular and flexible way. This makes it possible
to evaluate how different aspects of the biological-, reporter- and
detection- system influence the overall result of the experiment.

\subsection{Reactions and random motion in arbitrary geometries}
The microscopy simulation part of SMeagol uses input trajectories in
the form of a list of times, positions, particle id numbers, and
diffusive states, and can also include the time for creation and
destruction of particles. The input trajectories are interpolated
using Brownian bridges \cite{chow2009} to generate individual emission
positions of every simulated photon. Brownian bridges simulate free
diffusion, but the input data need not.  Thus, with a fine time step
one can use SMeagol to simulate general types of motion. \ML{For an
  example, see Sec.~\ref{SI5ex}b and Fig.~\ref{fig:helix}.}

We have extended the reaction-diffusion simulation software MesoRD
\cite{fange2012} to keep track of individual molecules and write
trajectories in the appropriate input format, and incorporated it in
SMeagol, but it is also possible to use indata from other
sources. SMeagol's trajectory data format is described in the software
manual.

\subsection{Tunable photophysics} 
In parallel with the diffusion process, each particle in the
simulation goes through a simulated stochastic photophysical process
which includes activation, Markovian transitions between multiple
photophysical states with different photon emission intensities, and
eventually irreversible photobleaching. Short exposures can be
simulated by setting an exposure time $t_E$ shorter than the frame
time $\Delta t$, and photophysical effects of excitation, by
specifying different photophysical transition rates during the
illuminated ($0\le t < t_E$) and non-illuminated ($t_E \le t<\Delta
t$) phases.  Separating photophysics and molecular diffusion makes it
possible to simulate the same reaction-motion trajectory under a wide
range of experimental conditions.

\subsection{EMCCD noise and background} 
The emitted photons are mapped to the camera chip using a point-spread
function (PSF) model, combined with simulated EMCCD \cite{ulbrich2007}
and background noise. The microscopy image, which is written to
tif-stacks for further analysis.

\subsection{Flexible, modular and user-friendly}
SMeagol is designed to allow easy incorporation of experimental data
and theoretical parameters at many levels.  Thus, the user can specify
arbitrary fluorophore activation and photophysical kinetics, and also
incorporate custom-written Matlab routines for PSF and background
models, by extending existing template files.  It is also possible to
use the independently measured PSF for a specific optical set-up, or
background movies from a specific sample.  Stochastic
reaction-diffusion models are described using the systems biology
markup language (SBML)\cite{finney2003} with extensions to spatial
models\cite{fange2012}. The trajectories and the different building
blocks of the microscopy simulation are then combined and
parameterized using either a graphical user interface, parameter text
files (runinput files), or Matlab structs.

\section{Point-spread function (PSF) model}
For all microscopy simulations described here, we used a rotationally
symmetric PSF model constructed from the Gibson-Lanni
model\cite{gibson1992}, as implemented in PSFgenerator
\cite{kirshner2013}.

We simulated the Gibson-Lanni PSF model with high resolution for 584
nm light, NA=1.4, and otherwise default settings PSFgenerator,
computed the cumulative radial distribution function (CRDF) for
different focal planes and radii up to 5 \si{\micro\metre}, and
constructed a Matlab look-up table for the inverse CRDF. An individual
photon emitted at $x_\text{em.},y_\text{em.},z_\text{em.}$ were then simulated as
detected at position
    \begin{IEEEeqnarray}{rcl}
      x_\text{det.}&=&x_\text{em.}+r\cos\nu,\\
      y_\text{det.}&=&y_\text{em.}+r\sin\nu,
    \end{IEEEeqnarray}
where the angle $\nu$ is uniformly distributed on $[0,2\pi]$, and $r$
is sampled using the inverse transform method \cite{press1992}, i.e.,
\begin{equation}
  r=CRDF^{-1}(u;z_\text{em.}),
\end{equation}
with $u$ uniformly distributed on $(0,1)$.  The total intensity of the
spot did not vary significantly in the region ($|z_\text{em.}-z_\text{focus}| <
400\,\si{\nano\metre}$) relevant for our simulations.  

In focus, the above PSF model has a standard deviation of about 335
nm. This is largely due to large shoulders of the PSF, and the width
of the central peak is about 95 \si{\nano\metre}.

\section{Simulated experiments with Min oscillations} 
To generate Fig.~1c, we simulated a minimal stochastic model of the
Min oscillation cycle\cite{huang2003,fange2006}, implemented using
scale-dependent mesoscopic reaction rate constants\cite{fange2010}.

\subsection{Reaction-diffusion simulation} 
To match experimental microscopy data, we choose an
\textit{E.~coli}-like geometry consisting of a cylinder with spherical
end caps, with outer diameter 1 \si{\micro\meter} and total length 2.7
\si{\micro\meter}. The outer 35 \si{\nano\meter} of the cell was
modeled as the membrane region. In this geometry, the oscillations
have a period of 25-30 s.

We used a 10 \si{\nano\meter} spatial discretization, and initial
conditions placing 1716 MinD and 483 MinE randomly in the membrane
region at one half of the cell.  After a 300 s simulation to reach
steady state, we ran a tracking production run, collecting snapshots
(used for the left column of Fig 1c) every 0.25 s, and tracking
positions of all species involving MinE every 5 \si{\milli\second}.

\subsection{MinE SPT simulation} 
(Fig 1c, mid columns, and supplementary movie S1). We simulated a
single particle tracking experiment with 4 ms frame rate and
continuous illumination. The coordinate system of the input
trajectories where rotated and translated to fit a brightfield image
of an E coli bacterium expressing no fluorophores (Fig.~1c), and a
fluorescence time-lapse movie of the same cell was used as a
background.

The Gibson-Lanni PSF model was used as described above, and an
experimentally parameterized blink/bleach model for
mEos2\cite{lee2012}, with a bright state intensity of 125000 photons/s
that yields on average 500 photons per spot and 4 ms frame if no
blinking occurs. Photoactivation events where simulated every 10 s,
with an activation probability of 20\% per unconverted molecule.

We simulated an EMCCD gain of 40 (SMeagol inverse gain
\texttt{camera.alpha=1/40}), i.e., every photon generates an
exponentially distributed number of image counts with mean value
40. Offset and readout noise are in this case included in the
experimental background.

\subsection{MinE long-exposure simulation} 
(Fig 1c, right column, and supplementary movie S2). For the
long-exposure simulations, we randomly activated 50\% of the
MinE-containing molecules with a constant emission intensity of 60
photons/s and no blinking or bleaching. For background, we set a
uniform intensity of 1 photon/pixel. We used continuous illumination
and a frame rate of 1 Hz, the Gibson-Lanni PSF model described above,
an EM gain of 40, offset 100, and readout noise with std.~4.

The reaction-diffusion model and trajectory output files, and the
SMeagol runinput files for these simulations, are included in
Supplementary dataset S1.

\section{Analysis of diffusion with blur and localization errors}
\begin{figure}
\begin{center} 
\includegraphics{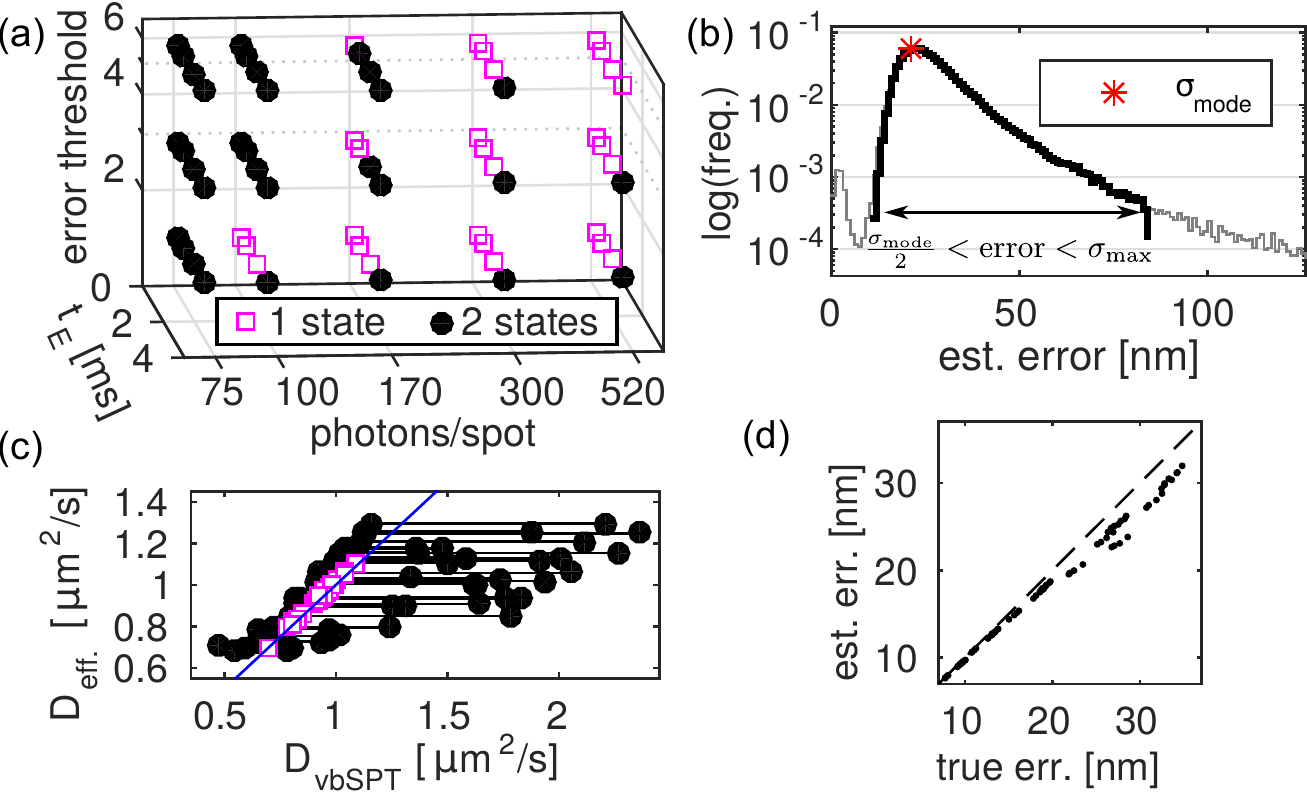}
\end{center}
  \caption{\label{fig2} Sensitivity of vbSPT to imaging artifacts.
    (a) Depending on the experimental parameters, vbSPT correctly
    finds one diffusive state (pink) or incorrectly finds two
    diffusive states (black). The experimental parameters are spot
    intensity, exposure times $t_E$, and localization error
    threshold. (b) Spot selection criteria illustrated on a
    distribution of estimated pointwise localization errors, with
    $\sigma_\text{max}/\sigma_\text{mode}$ being the error threshold
    in (a).  (c) Comparison of the diffusion constants found by vbSPT
    with an effective diffusion constant defined via the theoretical
    step length variance\cite{berglund2010}.  (d) Comparison of
      the true and estimated (as in (b)) root-mean-square error for
      every trajectory in (a).
%    Comparison of pointwise error estimates (root-mean-square averages
%    for every trajectory in (a)). Plugging point localization fit
%    parameters into the CRLB formula\cite{mortensen2010} ("CRLBfp") is
%    biased to overestimate the errors in the limit of many photons
%    (low error), while the likelihood-based estimate used in (b)
%    ("Laplace", Eq.~\eqref{eq:laplace}) is unbiased in this limit.
}
\end{figure}

To illustrate how SMeagol could be used to evaluate analysis methods
for live cell single particle tracking experiments, we explore the
ability of the vbSPT software\cite{persson2013} to correctly identify
the number of diffusive states in SPT data. We simulated an SPT
experiment with normal diffusion at a rate of
$D=1\;\si{\square\micro\meter\per\second}$ sampled every 3 ms in an
\emph{E.~coli} geometry with varying fluorophore brightness and
exposure time (see Supplementary movie S3). vbSPT assumes that data
come from a Markov model with state-dependent diffusion constants,
i.e., a model that neglects, e.g., z-dependent localization
errors\cite{deschout2012}, motional
blur\cite{deschout2012,berglund2010}, and confining effects of the cell
boundaries. The simulation thus contains many real features not
included in the analysis model, which could lead vbSPT to overfit the
data by incorrectly identifying more than one diffusive state. The
question is under which experimental conditions this is likely to
happen.

\subsection{Reaction-diffusion simulation}
We simulated simple diffusion of a single fluorescent particle in an E
coli-like geometry, built as a cylinder with length 3
\si{\micro\metre} and diameter of 0.8 \si{\micro\metre}, plus
spherical end caps. We used 10 \si{\nano\metre} voxels, and wrote
particle positions every 7 \si{\milli\second}.

\subsection{SMeagol simulation}
We generated SPT movies with a frame duration of 3 \si{\milli\second}
in a range of imaging conditions from the above diffusive trajectory,
by varying the exposure time $t_E$ and the average number of photons
per spot ($N_{phot.}=t_E\times \text{emission intensity}$). In
particular, we used all combinations of $t_E=
0.5\,\si{\milli\second}$, 1 \si{\milli\second}, 2\,\si{\milli\second},
3\,\si{\milli\second}, and $N_{phot.}=75$, 100, 300, 520.  Other
microscopy parameters used in all cases are summarized in table
\ref{tab:SMsettings_fig2}.
\begin{table}
  \begin{center}
  \caption{\label{tab:SMsettings_fig2} Settings for the microscopy
    simulations of simple diffusion (Fig.~\ref{fig2}).}
    \begin{tabular}{l@{ : }l}
      \hline
      sample time & 3 \si{\milli\second} per frame\\ 
      photophysics& constant emission intensity, no bleaching\\
      ROI & 80 nm pixels, $55\times20$ pixel ROI, focal plane in the
      mid-\\&plane of the bacteria\\
      camera & offset=100, readout noise std.=4, \\&
      EM gain=20 counts/photon\\
      background & constant, on average 1 photon/pixel per frame\\
      \hline
    \end{tabular}
  \end{center}
\end{table}

\subsection{Estimated number of states}
Fig.~\ref{fig2}a shows number of states learned by vbSPT as a function
of three tuning parameters: the exposure time $t_E$, the average
number of photons per spot, and the maximally allowed pointwise
localization error (Fig.~\ref{fig2}b).  The correct and overfitting
conditions are indicated in purple and black, respectively. In
general, all three parameters influence the overfitting tendency in a
non-trivial way. Continuous illumination (exposure time=frame time)
leads to overfitting in almost all conditions, but a modest decrease
in exposure time using, e.g., stroboscopic illumination\cite{elf2007},
leads to significant improvement due to decreased motional blur. We
also note that if the number of photons per localized molecule is
limited, it is advantageous to include only positions with high
localization accuracy.

%In the rest of this section, we give further details about simulation
%and analysis methods used in this simulated experiment.

\subsection{Estimating the diffusion constant}
As the analysis model of vbSPT neglects both localization errors and
motional blur, one should not take the numerical estimates of the
diffusion constants at face value. However, the estimates can be
interpreted using a theory of motional blur for diffusing
particles\cite{berglund2010}.  A closer inspection of the analysis
algorithm\cite{persson2013} shows that vbSPT effectively looks at the
step length variance, which in the absence of localization error and
blur is simply
\begin{equation}
  \mean{\Delta x^2}=2D_\text{vbSPT}\Delta t.
\end{equation}
A more detailed model that includes motional blur and localization
errors\cite{berglund2010} instead predicts
\begin{equation}
    \mean{\Delta x^2}=2D\Delta t(1-2R)+2\mean{\sigma_x^2},
\end{equation}
where $R=\frac{t_E}{3\Delta t}$ is the motional blur coefficient, and
$\mean{\sigma_x^2}$ is the mean-square localization error.
Eliminating the step length variance from the above equations, we
find that
\begin{equation}\label{eq:Deff}
  D_\text{vbSPT}=D_\text{eff.}\equiv D(1-2R)+\frac{\mean{\sigma_x^2}}{\Delta t}.
\end{equation}
Fig.~\ref{fig2}c plots the effective diffusion constant
$D_\text{eff.}$ vs.~the posterior mean of $D_\text{vbSPT}$ for the
different data sets (using our estimated average localization errors),
and we see that the prediction of Eq.~\eqref{eq:Deff} is reproduced
well when a single diffusive state is correctly identified.

%Independent of parameter setting, vbSPT correctly estimates an
%effective diffusion constant corresponding to the step
%variance\cite{persson2013}. Thus, we can use the theory of motional
%blur\cite{berglund2010} to define an effective diffusion constant
%given by
%\begin{equation}
%2D_\text{eff}\Delta t\equiv \mean{\Delta x^2}
%=2D\Delta t(1-2R)+2\mean{\sigma_x^2},
%\end{equation}
%where $\Delta t$ is the inverse frame rate, $R=\frac{t_E}{6\Delta t}$
%is the motional blur coefficient\cite{berglund2010}, and
%$\mean{\sigma_x^2}$ is the variance of the localization error. As
%shown in Fig.~\ref{fig2}c, vbSPT and our estimated errors reproduce
%this prediction very well when a single diffusive state is correctly
%identified.

\subsection{Point localization}
We localized the spots using a maximum-likelihood fit of a symmetric
Gaussian plus constant background to a 7-by-7 fit region, using the
EMCCD likelihood function of Ref.~\citenum{mortensen2010}, with the
offset and gain settings of table \ref{tab:SMsettings_fig2}. Each spot
is thus described by 5 fit parameters: background $b$, spot amplitude
$N$, spot standard deviation $s$, and spot position $\mu_x,\mu_y$. We
used Matlab's built-in function \texttt{fminunc} for numerical
optimization of the log likelihood, which was parameterized to allow
only positive values of $b$, $N$, and $s^2$, and used the true spot
positions, and the average PSF width and amplitude to construct an
initial guess for each fit.  To minimize confinement artifacts from
the cell walls, we analyzed motion and uncertainties along the long
cell axis ($x$ coordinate) only.
\subsection{Estimating point-wise localization uncertainty}
Due to fluorophore motion during exposure, random photon emission,
$z$-dependence of the PSF, etc., the quality of the fit varies from
spot to spot. We used a Laplace approximation\cite{mackay2003} (also
known as the saddle point approximation in statistical physics) of the
likelihood function to estimate the localization uncertainty of
individual spots, as follows: Let $IM$ denote the fit region of the
image used for localization, and $\vec\theta=(\mu_x,\mu_y,\ldots)$ the
fit parameters. We approximate the likelihood function
$p(IM|\vec\theta)$ by a Gaussian centered at the maximum likelihood
estimate $\vec\theta^*$ using a Taylor expansion in
$(\vec\theta-\vec\theta^*)$,
\begin{multline}\label{eq:saddlept}
  p(IM|\vec\theta)
  %=e^{\ln p(IM|\vec\theta)}
  \approx \exp\Big(
  \ln p(IM|\vec\theta^*)
  +\underbrace{\nabla_{\vec\theta}\ln P(IM|\vec\theta)\big|_{\vec\theta^*}}_{=0}
  (\vec\theta-\vec\theta^*)\\
  -\frac 12 (\vec\theta-\vec\theta^*)^T\Sigma^{-1}(\vec\theta-\vec\theta^*)
  +\ldots
  \Big),
\end{multline}
where the first order term disappears since $\vec\theta^*$ is a local
maximum, and the second order term is given by the inverse covariance
matrix,
\begin{equation}\label{eq:laplace}
  \Sigma^{-1}=-\frac{\partial^2\ln p(IM|\vec\theta)}{\partial\vec\theta^2}
  \bigg|_{\vec\theta^*}.
\end{equation}
This can be interpreted as the Bayesian posterior distribution (with a
flat prior). The uncertainty of the parameters are then characterized
by their posterior covariances\cite{mackay2003}. In particular, the
posterior variance of $\mu_x$ is approximately given by
$\Sigma_{\mu_x,\mu_x}$.

As a simple test of this estimator, we compare the true and
estimated average root-mean-square (RMS) error for all points in
every trajectory (Fig.~\ref{fig2}d). We find it to be correct on
average, i.e.,
\begin{equation}
  \mean{(\mu_x^*-\mu_{x,\text{true}})^2} \approx \mean{\Sigma_{\mu_x,\mu_x}},
\end{equation}
for true RMS errors $\lesssim 20$ nm, but biased downwards for
larger errors, probably because the Gaussian approximation of the
posterior density (Eq.~\eqref{eq:saddlept}) is inaccurate in those
cases.

\subsection{Selection criteria}
To build diffusion trajectories for vbSPT analysis, we first discarded
spots where the numerical optimization failed. We then built a
histogram of estimated standard errors
$\sigma_x=\sqrt{\Sigma_{\mu_x,\mu_x}}$, and identified the most likely
estimated error $\sigma_\text{mode}$.  One such histogram is shown in
Fig.~\ref{fig2}b.  Finally, we discarded spots that had either
$\sigma_x<\sigma_\text{mode}/2$ as being unrealistically precise, or
$\sigma_x> \sigma_\text{mode} \times\text{(error threshold)}$, using
error thresholds in the range $1.3 -6$ as our third control variable
in Fig.~\ref{fig2}a. The highest threshold of 6 included practically
all spots.

The fraction of retained spots as well as average trajectory length
vary with both simulation parameters and error threshold, but for this
experiment we generated enough images to construct data sets with 80
000 diffusive steps for all conditions.

\subsection{vbSPT settings}
For the trajectory analysis, we used vbSPT 1.1.2\cite{persson2013}.
For each data set, we ran 20 independent runs of the greedy model
search algorithm with up to 15 hidden states.  We used an inverse
gamma prior with mean value 1 \si{\square\micro\metre\per\second} and
strength 5 (std.~$\approx 0.6$ \si{\square\micro\metre\per\second})
for the diffusion constants (with 80 000 diffusive steps in the
trajectory, this prior is completely overwhelmed by the data), flat
Dirichlet priors for the initial state and state-change probability
distributions, and a Beta distribution with mean 0.02 s and std.~2 s
for the mean dwell time of the hidden states. (For detailed
definitions, we refer to the vbSPT manual).

\section{Misc.~model examples}\label{SI5ex}
Here, we briefly describe some additional model examples, to illustrate
the capability and limitations of SMeagol in various settings. Source
files for these examples are included in Supplementary data S5.
\begin{figure}
  \includegraphics{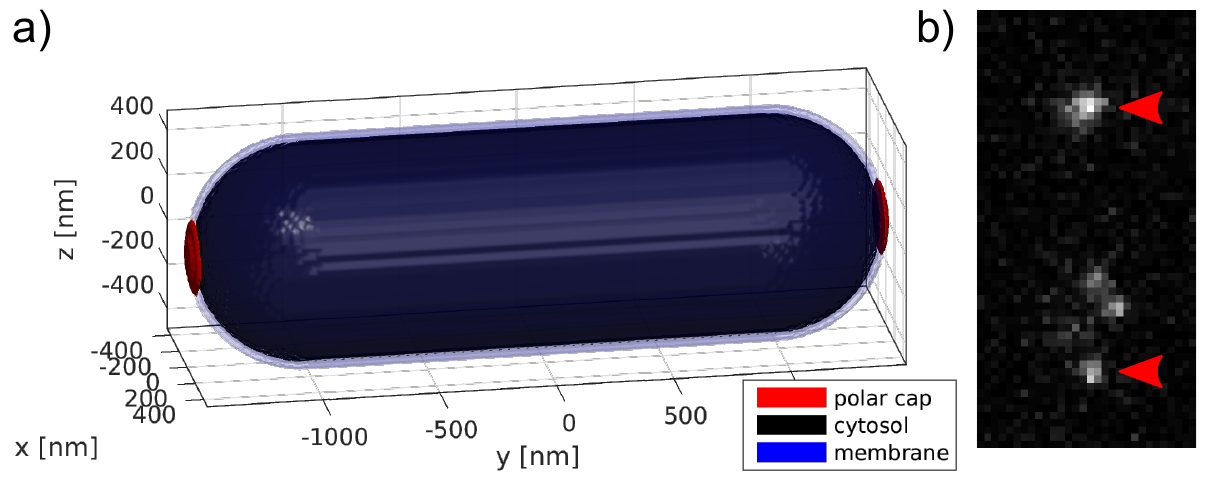}
  \caption{\label{fig:popz} Simulating a diffusing protein with polar
    binding regions. a) Model geometry, with the cytosol (black)
    inside a thin membrane region (blue) that also contains polar caps
    (red,magenta) where binding receptors are localized. b) Snapshot
    from a simulated SPT experiment, with both polar bound (red
    arrows) and freely diffusing molecules visible.}
\end{figure}
\subsection{\ref{SI5ex}a. Affinity to cell poles}
To illustrate the geometric modeling capabilities of MesoRD and
SMeagol, we construct a simplified model of a diffusing cytoplasmic
protein P with specific affinity to binding partners localized at the
cell poles.

We use an E coli-like rod shape, where cytoplasm is modeled by a union
of a cylinder and two spheres with radii 465 nm. In the cytoplasm, the
proteins diffuse with diffusion constant 2.5
\mbox{\si{\square\micro\metre\per\second}}.  The membrane region is
modeled as an additional 35 nm outer shell, from which polar regions
in the form of small spherical caps are created, as shown in
Fig.~\ref{fig:popz}a. In these caps, we assume a constant
concentration [R] of receptors. Then, the binding rate can be written
$r=k_a[R][P]$, where $k_a$ is the association rate constant, and the
unbinding rate constant is $k_d$. We choose $k_a[R]=20$
\si{\per\second}, $k_d=0.01$ \si{\per\second}, and diffusion constant
0.01 \si{\square\micro\metre\per\second} for the bound complex
(confined to the membrane caps).

We ran 6 s of stochastic reaction-diffusion simulations starting with
125 P molecules uniformly distributed outside the polar caps, and
after a burn-in of 2 s, wrote positions to file every 1
\si{\milli\second}.  For the microscopy simulation, we randomly
activated 5\% of the molecules, and used the same settings as in the
simple diffusion experiment above (table \ref{tab:SMsettings_fig2}),
except slightly larger region of interest, a fluorophore intensity
corresponding to giving on average 270 photons/frame and fluorophore,
and photobleaching with a mean lifetime of 1 s. A snapshot is shown in
Fig.~\ref{fig:popz}b.

\begin{figure}
  \includegraphics{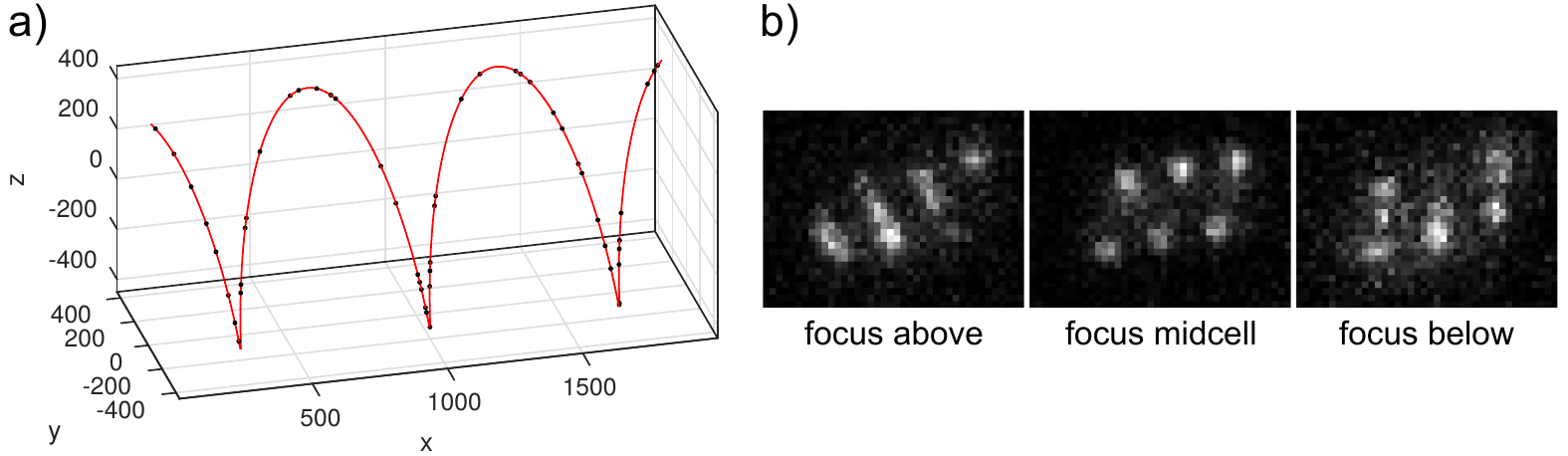}
  \caption{\label{fig:helix} Simulating non-diffusive transport. (a)
    Snapshot from TASEP simulation (black, particles) on a helical
    path (red). (b) Simulated images with focal plane placed near the
    upper, middle, and lower part of the helical curve emphasize
    different parts of the path.}
\end{figure}
\subsection{\ref{SI5ex}b. 
Non-diffusive transport along helical membrane filaments} To
illustrate the possibility of simulating more complex motion than
diffusion using SMeagol, we constructed an active transport model of
particles moving on a helical path (Fig.~\ref{fig:helix}). More
specifically, we wrote a Matlab script to simulate a totally
asymmetric exclusion process (TASEP\cite{lazarescu2011}) with open
boundaries: Particles (black) are created at the left end of the helix
in Fig.~\ref{fig:helix}a, walk forwards in 36 nm steps along the
helical path (red) with a rate of 10 \si{\per\second} (under a site
exclusion constraint), until they fall off at the right end. The
insertion rate was 2 \si{\per\second}, putting the TASEP in the low
density phase \cite{lazarescu2011}. We then wrote particle coordinates
to a trajectory text file at regular intervals (all in the same
chemical state), and fall-off events as particle destructions in the
reactions text file. For microscopy simulations, we set $D=0$, which
disables the Brownian bridges and leads to linear interpolation
between the trajectory coordinates.  Simulation scripts and runinput
files are included in data set S5.

\begin{figure}
  \includegraphics{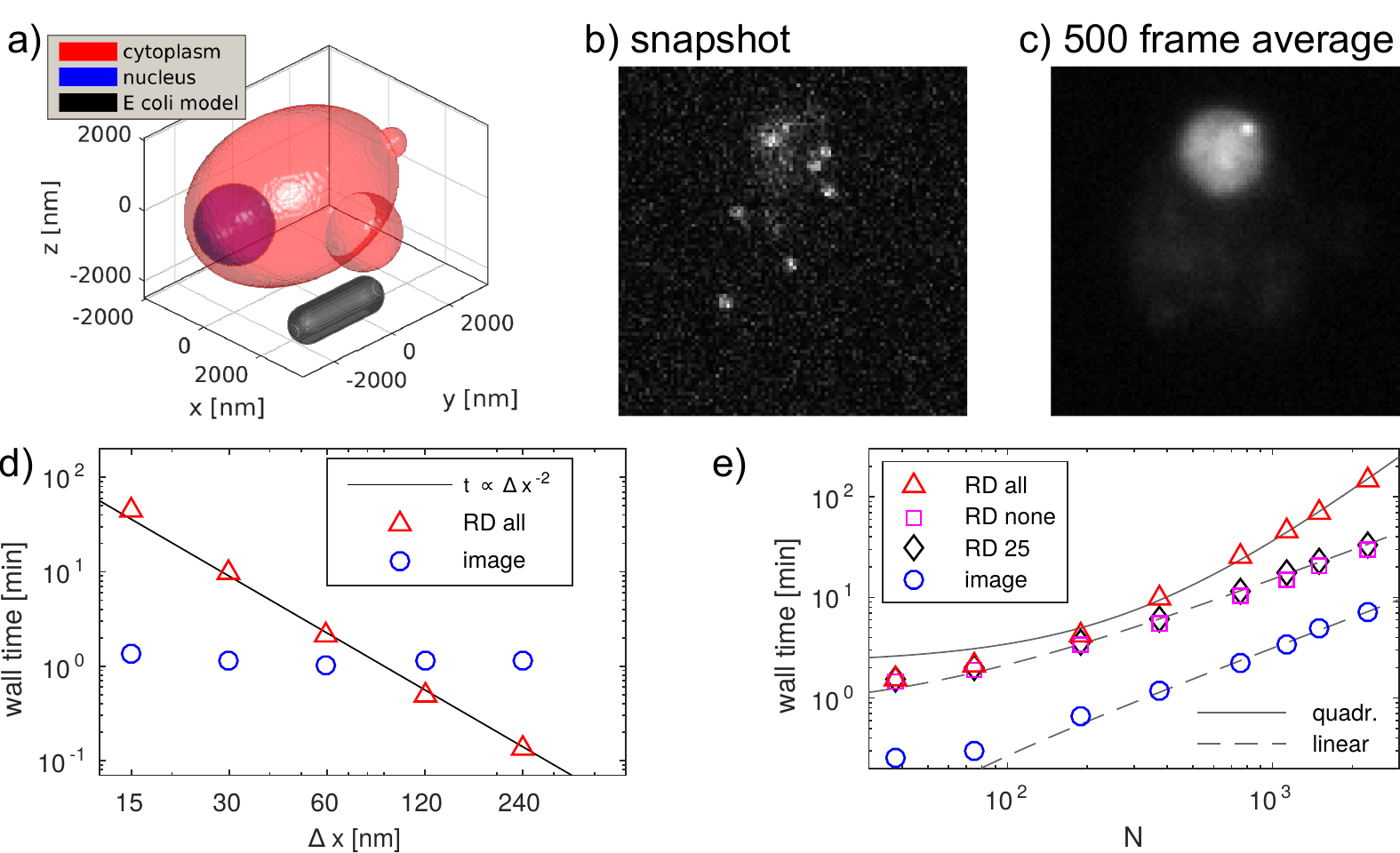}
  \caption{\label{fig:yeast} Simulating a diffusing transcription
    factor in a yeast cell. (a) The model geometry includes a
    cytoplasm compartment with two buds (red) and a spherical nucleus
    (blue). Also shown is the bacterial model of Fig.~\ref{fig:popz}
    (black). (b,c) A snapshot and 500-frame average, respectively,
    from a 5 ms/frame SPT simulation. (d) Wall time versus voxel size
    for a 5 s stochastic reaction diffusion simulation tracking all
    proteins (RD all), and microscopy image simulation. The former
    closely follows the expected $1/\Delta x^2$ scaling, while the
    latter is essentially constant. (e) Wall times for the $\Delta
    x=30$ nm case from (d), but with varying number of involved
    proteins. ``RD none'' and ``RD 25'' refers to stochastic reaction
    diffusion simulations with tracking deactivated and tracking only
    25 proteins irrespective of total copy number. }
\end{figure}

\subsection{\ref{SI5ex}c. A large cell}
To illustrate the computational requirements of SMeagol, we use a
simple model of transcription factor (TF) motion in a budding yeast
cell. The model has a cytoplasmic compartment, made of two spheres
joined to an ellipse with a spherical nucleus compartment inside. As
seen in Fig.~\ref{fig:yeast}a, it is several times larger than the
bacterial-like examples described so far.

The main computational load for fine-scale MesoRD simulations consists
of diffusive motion (total diffusive hopping rate is $6D/\Delta x^2$),
so we use a simple kinetic model where the TF has a 'free' state that
diffuses with diffusion constant 2.5
\si{\square\micro\metre\per\second} in the entire cell, and can
interconvert to a 'bound' state (0.01
\si{\square\micro\metre\per\second}) inside the nucleus. In addition,
we simulate active transport into the nucleus by setting the diffusion
rate from the cytoplasm to the nucleus 20 times larger than that in
the reverse direction. Protein accumulation in the nucleus is clearly
visible in the simulated images (Fig.~\ref{fig:yeast}b,c).

Fig.~\ref{fig:yeast}d shows wall time\footnote{All computing times
  were measured on a 2.40GHz Intel Xeon CPU with 24 GB RAM. } for a 5
s stochastic reaction-diffusion simulation, starting with about 375
uniformly distributed TFs and various voxel sizes, which fits very
well with the expected inverse quadratic scaling $\propto \Delta
x^{-2}$. Second, we generated a 2.5 s simulated microscopy movie with
5 ms frame time, $100\time 100$ 80 nm pixels, and 10\% of the proteins
activated and emitting on average 270 photons/frame. This simulation
is limited by evaluating Brownian bridges and parsing the trajectory
file, and thus independent of the discretization of the trajectories.

Secondly, we varied the number of proteins, while keeping the
discretization constant at $\Delta x=30$ nm. As seen in
Fig.~\ref{fig:yeast}d, the image simulation part scales linearly as
expected, while the RD simulation scales quadratically if all proteins
are tracked, but linearly with tracking deactivated or if tracking only
a fixed number of proteins.  Thus, we see that while fairly large
cells can be simulated, the RD simulations can be computationally
demanding especially for models that include large cell size, tracking
of many molecules, and small details that require fine
discretizations.

Finally, we note that the total size in itself does not influence
computing time significantly, as long as the number of subvolumes fit
in the RAM memory of the computer.  As an illustration, we rescaled
all length in the $\Delta x=30$ nm case of Fig.~\ref{fig:yeast}d by a
factor 0.3 (thus decreasing the volume by 97\%) but kept the
discretization and number of molecules constant. However, the RD
simulation of the smaller model only took 25\% less time.

\begin{figure}
  \includegraphics{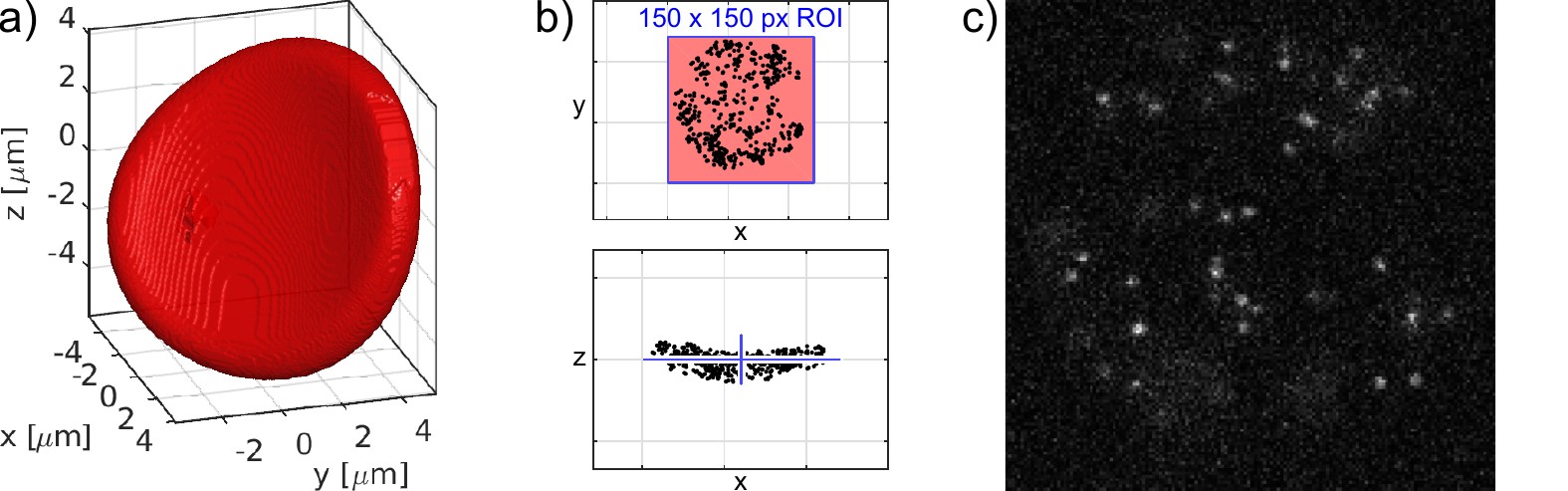}
  \caption{\label{fig:bloodcell} Modeling an erythrocyte. a)
    Rendering of the SBML model with 50 nm subvolumes. b) Alignment of
    the model (particle positions indicated by black dots) relative to
    the simulated region of interest (ROI) and focal plane of the
    simulation. c) Snapshot from a simulated movie with 25 ms exposure
    time.}
\end{figure}
\subsection{\ref{SI5ex}d. Complex shapes}
To highlight SMeagol's abilities to model geometries beyond the
tractable possibilities given by transformation and combinations of
simple geometric primitives, we also model particle diffusion in an
erythrocyte shaped geometry, shown in Fig.~\ref{fig:bloodcell}a. The
starting point is here a freely available 3D model\footnote{
  \url{http://www.turbosquid.com/FullPreview/Index.cfm/ID/509576},
  accessed 2016-02-04)}, which we imported to and exported from
Blender\footnote{\url{www.blender.org}} to make it a triangular
mesh. The mesh was then converted to SBML compatible format using a
custom Python script, and incorporated into an SBML model, where about
70 particles diffuse freely (D=2 \si{\square\micro\metre\per\second})
within the erythrocyte.

For the microscopy simulation, we rotated the model to align it with
the focal plane as shown in Fig.~\ref{fig:bloodcell}b and used similar
settings as for the yeast example to produce the snapshot in
Fig.~\ref{fig:bloodcell}c.

\section{Misc.~supplementary material}
\begin{itemize}
\item {\bf Supplementary movie S1} illustrates the simulated MinE
  single particle tracking experiment (Fig.~1c, mid columns), showing
  both true particle positions, the experimental background, and the
  simulated result in three separate panels.
\item {\bf Supplementary movie S2} illustrates the simulated MinE
  fluorescence microscopy time-lapse movie (Fig.~1c, right column), as
  well as the true particle positions.
\item {\bf Supplementary movie S3} shows simulated single particle
  tracking experiments that are analyzed in Fig.~\ref{fig2}.
\item {\bf Supplementary dataset S4} contains the SBML files,
  reaction-diffusion trajectory output, and SMeagol runinput files for
  the simulated Min oscillation experiments (Fig.~1c).
\item {\bf Supplementary dataset S5} contains model and runinput
  files, scripts, and source files for the example models described in
  Sec.~\ref{SI5ex}.
\end{itemize}

%%%% % hack: to continue enumerating starting with ref 21, 
%%%% % 1. create mainrefs.bbl and SIrefs.bbl from SMeagol_SI.bbl
%%%% % 2. add the following line to SIrefs.bbl just before the first bibitem:
%%%% % \setcounter{NAT@ctr}{20}
%%%% % also change the bibliography heading
%%%% \renewcommand{\refname}{Supplementary references}
%%%% 
\end{document}